\documentclass[a4paper,10pt]{article}
\usepackage{amsmath}
\title{ Odd Hamiltonian Structure for Supersymmetric Sawada - Kotera Equation }

\author{Ziemowit Popowicz \\
        Institute of Theoretical Physics
        University of Wroc{\l}aw \\
        pl. M. Borna 9
        Wrocław Poland \\
        ziemek@ift.uni.wroc.pl\\}

\begin{document}

\maketitle

\begin{abstract}
We study the  supersymmetric $N=1$ hierarchy connected with the
Lax operator of the supersymmetric  Sawada-Kotera equation. This operator
produces the physical equations as well as the exotic equations with odd time.
The odd Bi-Hamiltonian structure for the N=1 Supersymmetric Sawada - Kotera equation is defined.
The product of the symplectic and implectic Hamiltonian operator gives us the recursion operator.
In that way we prove the integrability of  the supersymmetric Sawada - Kotera equation in the sense that it has the Bi-Hamiltonian structure. The so called ``quadratic'' Hamiltonian operator of  even order generates the  exotic equations while the  ``cubic'' odd Hamiltonian operator generates the physical equations.

\end{abstract}

\vspace{2cm}

\section{Introduction}

Integrable Hamiltonian systems plays an important place in diverse branches of theoretical physics as exactly solvable models of fundamental physical phenomena ranging from nonlinear hydrodynamics to string theory \cite{Fad}. There are several different approaches which  generalize these models.

One of them is to extend the theory by addition of fermion fields.
The first results in this direction can be found in {\cite{Kuper0,Masud,Auria,Gurses,Kulish,Manin}}. 
In many cases, the extension by fermion fields does not guarantee the supersymmetry of the final theory. Therefore this method was called the fermion extension in order to distinguish it from the
fully supersymmetric method which was developed later {\cite{Mat1,Mat2,Mat3,Masud1}}.

There are many prescriptions how to embed a given classical  model into a  fully supersymmetric one.
The main idea is simple: in order to obtain an $N$ extended supersymmetry   multiplet
 one has  to add to a system of $k$ boson fields
$kN$ fermions and $k(N-1)$ bosons $k=1,2,\dots, N=1,2,\dots$.
Working with this supermultiplet we can  apply  integrable Hamiltonians methods.

From the soliton theory point of view we can distinguish two important classes of
the supersymmetric equations: the non-extended $N=1$ and extended $N>1$ cases.
The extended case may imply new bosonic equations whose properties need further investigations.
Interestingly enough, some typical  supersymmetric effects appearing during the supersymmetrizations
have important consequences for the soliton theory.
Let us mention for instance: the nonuniqueness of the roots of the supersymmetric Lax operator \cite{Pop1},
the lack of a bosonic reduction to the classical equations \cite{Ivan},
occurrence of non-local conservation laws {\cite{Kersten, Mat4}}
and existence  of odd Hamiltonian structures {\cite{Bek, Pop2,Das}}.

The odd Hamiltonian structures  first  appeared in {\cite{Butin}}. It was latter  noticed in
{\cite{Lejt}} that in the superspace one can consider both even and odd symplectic structures, 
with even and odd Poisson brackets, respectively. Recently the odd brackets, also known as
antibrackets or Buttin bracket, have been extensively investigated
 both from  the mathematical and the physical point of view.
They have drown some interest in the context of BRST formalism {\cite{Batalin}}, in the supersymmetric 
quantum mechanics {\cite{Soroka}}, in the classical mechanics {\cite{Kuper,Fryd,Khud,Soroka1}} and in the gravity theory \cite{Figa}. Becker and Becker {\cite{Bek} were probably the first who
discovered  odd Hamiltonian structures in the special non-extended
supersymmetric KdV equation. Later these structures have been discovered in many supersymmetric integrable models {\cite{Pop2,Das}}. This nonextended supersymmetrization, sometimes
called the $B$ extension, are based  on  the simple transformation of
dependent variables  describing the classical system,  onto superbosons. 
In that manner  the equations are not changed but are rather rewritten in theses new coordinates in the superspace. Such supersymmetrization is however ``without the supersymmetry'' because the bosonic sector remains the same and does not feel any changes after  supersymmetrization.

In this paper we found new  unusual features of the supersymmetric models.
First we show that recently discovered {\cite{Liu}} $N=1$ supersymmetric  extension of the Sawada - Kotera equation, (which is not a $B$ extension)  has an odd Bi-Hamiltonian structure. Probably it is a first nontrivial supersymmetric model with  an odd Hamiltonian structure  and a  modified bosonic sector.
The conserved quantities are constructed from the fermionic and bosonic fields
and disappear in the bosonic or  in the fermionic limit.
This example shows that from the superintegrability of the whole  system one can not conclude an integrability of the bosonic sector.

 The second observation is that Lax representation, which generates the superymmetric Sawada - Kotera equation,
 generates also   exotic equations with odd time. In this paper we treat these equations as subsidiary equations. This  allowed us to discovered the Bi-Hamiltonian structure for the supersymmetric Sawada-Kotera equation in which the  symplectic and the implectic operators are odd.
 In that way we proved the integrability of the model in the sense that it possess  the Bi-Hamiltonian structure.

The next observation is that the usual even supersymmetric second Hamiltonian operator of the supersymmetric Korteweg - de Vries equation generates also  exotic  equations.  Unfortunately we have been not able to discovered the second Hamiltonian operator for these exotic  equations.

We also showed  that the odd  and the even Hamiltonian operator
 could be derived systematically using the so called $R$ matrix theory
{\cite{Hamy1,Hamy2,Hamy3,Hamy4,Hamy5,Hamy6}}.

The paper is organized as follows. In first section we recapitulate known facts about the supersymmetric and the $B$ extensions of
 Korteweg -  de Vries equation and its Bi-Hamiltonian structure. In second section
we describe the $N=1$  supersymmetric Lax representation   which produces  supersymmetric Sawada - Kotera,
 the $B$ extension of Kaup - Kupershmidt and the Sawada - Kotera equation. 
The third section contains our main results. We demonstrate   the odd Bi - Hamiltonian structure for the
 Sawada - Kotera,  which has been  accidentally discovered from the general considerations
and we define the even Hamiltonian operator for the exotic equations. Next we check
the validity of the Jacobi identity for our Hamiltonian operator.
Finally the proper choice of our  Hamiltonians structure is confirmed by
computing once more these structures using the so called classical $R$ -matrix theory.

\section{ Supersymmetric and B extension of the Korteweg - de Vries Equation}

Let us consider the  Korteweg - de Vries equation in the form
\begin{equation}
 u_x = u_{xxx} + 6uu_x
\end{equation}
where the function $u$ depends on $x$ and $t$. This equation is usually rewritten  in two different forms.
The first one utilizes the Bi-Hamiltonian formulation while in the second the Lax representation is used.
We have the following Bi-Hamiltonian formulation of the KdV equation
\begin{equation}\label{kdv}
 u_t = J \frac{\delta H_1}{\delta u} = \partial \frac{ \delta H_1}{\delta u} = K \frac{\delta H_2}{\delta u} = (\partial^{3} + 2\partial u + 2u \partial )
\frac{\delta H_2}{\delta u},
\end{equation}
where $H_1=\frac{1}{2}\int (uu_{xx} + 2u^{3}) dx$ and $H_2 =\frac{1}{2}\int u^{2} dx$ are conserved
quantities while $J=\partial_x$ and $ P=\partial_{xxx} + 2\partial _x u + 2u\partial_x$ constitute
the Hamiltonian operators.
The Lax representation is
\begin{equation}\label{lak}
 L_{t} = \big [ L, L_{+}^{\frac{3}{2}} \big ],
\end{equation}
where $L=\partial^{2} + u$.

We have two possibilities to expand the superfunction   $\varPhi $ as  $ w(x) + \theta\xi $ where $w(x)$ is
an even function  and  $\xi(x)$ is an odd function in the case of the superboson while in  the
case of superfermion as $\xi(x) + \theta w(x)$.
Hence we can represent the co-vector and the vector fields in two different ways.
For the superboson functions we choose the corresponding chart for the vector fields $K$  and co-vector
fields $\varUpsilon$  as
\begin{equation}\label{K}
 K(\varPhi) = K_{w} + \theta K_{\xi} \quad \quad \varUpsilon(\varPhi) = \varUpsilon_{\xi} +
\theta \varUpsilon_{w} \
\end{equation}
while for  superfermions as
\begin{equation}
 K(\varPhi) = K_{\xi} + \theta K_{w} \quad \quad \varUpsilon(\varPhi) = \varUpsilon_{w} -
\theta \varUpsilon_{\xi}
\end{equation}
Then the duality between the vector and the co-vector fields becomes the usual one
\begin{equation}
 \textless K  , \varUpsilon  \textgreater = \int K \varUpsilon dx d\theta = \int
(K_{w} \varUpsilon_{w} + K_{\xi} \varUpsilon_{\xi} ) \ dx
\end{equation}
and the conjugation becomes
\begin{equation}
 \textless K  , \varUpsilon  \textgreater ^{\star}  = \textless \varUpsilon  , K  \textgreater =
\int (K_{w} \varUpsilon_{w} +  K_{\xi} \varUpsilon_{\xi} ) \ dx = \textless K  , \varUpsilon  \textgreater
\end{equation}

Let us now try to find the supersymmetric $N=1$ extension of the KdV equation considering  the most general form of the Lax operator
\begin{equation}
 L=\partial^2 + \lambda_1 \Phi_{1}  + \lambda_2 \Phi {\cal D}
\end{equation}
where $\lambda_1$ and $\lambda_2$ are arbitrary constants, $\Phi $ is a superfermionic field of the
fractional conformal dimension $\frac{3}{2}$  with the
decomposition  $ \Phi = \xi + \theta u $ where  $ \xi $ is  an odd function and $u$ is an  even function.
${\cal D} = \partial_{\theta} + \theta \partial $ is the supersymmetric derivative with the property
$ {\cal D}^{2}= \partial$. In the next we use the following notation:
 $\Phi_{n}$ denotes the value of the action of the operator ${\cal D}^{n}$ on the superfunction $\Phi$,
 for example $\Phi_{0} = \Phi, \Phi_{1} =({\cal D}\Phi),\Phi_{2} = \Phi_{x}$ while
 ${\cal D}\varPsi = \varPsi_{1}  (-1)^{[\varPsi]} \varPsi {\cal D}$
 where $[\varPsi]$ denotes the parity of the superfunction $\varPsi$.

The consistency of  Lax representation (\ref{lak})  gives us  two solutions for the
coefficients $\lambda_1,\lambda_2$

The first solution $ \lambda_1=0,\lambda_2=1$ is the usual well known supersymmetric extension of the KdV equation
\begin{equation}
 \Phi_t= \Phi_{xxx} + 3(\Phi \Phi_{1})_{x}
\end{equation}
In the components this equation takes the form
\begin{eqnarray}
 u_t &=& u_{xxx} + 6uu_{x} - 3\xi\xi_{xx} \\ \
\xi_t &=& \xi_{xxx} +3(u\xi)_{x}
\end{eqnarray}

This supersymmetric $N=1$ KdV equation could be rewritten in the Bi-Hamiltonian form \cite{Pop1}
\begin{eqnarray} \label{skdv}
 \Phi_t & = & \varPi  \frac{\delta H_1}{\delta \Phi} =
  ({\cal D} \partial^{2} + 2 \partial  \Phi  +  2\Phi \partial  + {\cal D} \Phi  {\cal D})
\frac{\delta H_1}{\delta \Phi} \\
\Omega  \Phi_t & = & ({\cal D} \partial^{-1} + \partial^{-1}\Phi\partial^{-1})  \frac{\delta H_2}{\delta \Phi}
\end{eqnarray}
where
\begin{eqnarray}
 H_1&=&\int \frac{1}{2} \Phi \Phi_{1} dx d\theta = \int(\frac{1}{2} (u^{2} - \xi\xi_{x})\\
H_2&=&\int (\frac{1}{2} \Phi_{x}\Phi_{1,x} + \Phi\Phi_{1})^{2} dx d\theta =
\int \frac{1}{2}(u_{x}^{2} +2u^{3} + \xi_{xx}\xi_{x} + 4\xi_x\xi u) dx.
\end{eqnarray}

Regarding $\varPi$ operator as a linear map from the co-vector fields to the vector fields the
equation (\ref{skdv})  could  be decomposed into
\begin{equation}
\frac{d }{d t} \left(\begin{tabular}{cc}
$u$ \\ $\xi$
\end{tabular} \right) =
\left(\begin{tabular}{cc}
 $\partial^{3} +2\partial u + 2 u \partial $ & $ \partial \xi + 2 \xi \partial $ \\
$2\partial \xi + \xi \partial $ & $ -\partial^{2} - u $
\end{tabular}
\right) \left ( \begin{tabular}{c}
 $\frac{\delta H_1}{\delta u} $ \\
$ \frac{\delta H_1}{\delta \xi} $
\end{tabular}\right )
\end{equation}

Now regarding $\Omega$ as map from the vector fields to the co-vector fields we translate it to the matrix form
\begin{equation}
\Omega =
 \left( \begin{tabular}{cc}
$\partial^{-1}$ & $ \partial^{-1} \xi \partial^{-1}$ \\
$\partial^{-1} \xi \partial^{-1}$ & $-1 - \partial^{-1} u \partial^{-1} $
\end{tabular} \right ).
\end{equation}
\vspace{0.5cm}

The second solution  $ \lambda_1=1,\lambda_2=0$ leads us to the so called $B$ extension of the KdV equation
\begin{equation}\label{bbkdv}
 \Phi_t=\Phi_{xxx} + 6\Phi_{x} \Phi_{1}
\end{equation}
or in  the components
\begin{eqnarray}\label{bkdv}
 u_t &=& u_{xxx} + 6u_xu,\\ \
 \xi_t &=& \xi_{xxx} + 6\xi_xu
\end{eqnarray}
from which we see that this $B$ extension is ``without supersymmetry'' because the bosonic sector (\ref{bkdv})
coincides with  the Korteweg - de Vries equation. Moreover the  equation (\ref{bbkdv}) also coincides with
the usual KdV equation in which we simply replace the function $ u $
with the superfunction $\Phi_{1}$.   These procedure can be  considered as the intermediate
step between the KdV equation and the potential KdV equation. Indeed applying once more
the B-extension to the B-extension of the KdV equation, where now $\Phi = w_{1}$
and $w$ is a superboson function,  we obtain the potential KdV equation.

In general we can carry out such $B$ supersymmetrization to each equation of the form $u_t = F(u)_x$ and easily
obtain  its Bi-Hamiltonian structure, if such exists.
Indeed taking into account the transformation between $ u$ and $\Phi$ we can  obtain the
Bi-Hamiltonian structure for the B-extension of the KdV equation

\begin{equation}
\Phi_t={\cal D}^{-1} (\partial^{3} + 2(\partial \Phi_{1} + \Phi_{1} \partial)) {\cal D}^{-1}
\frac{\delta H_1}{\delta \Phi} = {\cal D}^{-1} \partial {\cal D}^{-1} \frac{\delta H_2}{\delta \Phi} =
\frac{\delta H_2}{\delta \Phi}
\end{equation}

where
\begin{eqnarray}
 H_1 & = & \frac{1}{2}\int dx d\theta \Phi \Phi_x = \int dx  \xi_xu \\ \
H_2 & = & \frac{1}{2} \int(\Phi\Phi_{xxx} + 2 \Phi_{1}^{3})) = \int dx ( \xi_{xxx}u + 3\xi_{x}u^{2})
\end{eqnarray}.
These equation could be translated to the matrix form

\begin{equation}
 \frac{d }{d t} \left(\begin{tabular}{c}
$u$ \\ $\xi$
\end{tabular} \right) =
\left(\begin{tabular}{cc}
 $ 0 $ &  $ -\partial^2 - 4u - 2u_{x}\partial^{-1} $ \\
$$ &  $$ \\
$\partial^2 + 4u -2\partial^{-1}u_x  $ &  $  -2\xi_{x}\partial^{-1} - 2\partial^{1}\xi_{x} $
\end{tabular}
\right)
\left ( \begin{tabular}{c}
$\frac{\delta H_1}{\delta u}$ \\
$$ \\
$\frac{\delta H_1}{\delta \xi}$
\end{tabular}\right ) = \left(
\begin{tabular}{cc}
$ 0  $  & $ -1 $ \\
$$ & $$ \\
$ 1 $ &  $ 0  $
\end{tabular} \right)
\left(\begin{tabular}{c}
 $ \frac{\delta H_2}{\delta u} $ \\
$$ \\
$ \frac{\delta H_2}{\delta \xi} $
\end{tabular}\right )
\end{equation}

There is a basic difference  between the second Hamiltonian operator of the supersymmetric extension and
the second Hamiltonian operator of  the B-extension of the KdV equation. The first one is an even operator while
the second is an odd one.
For further investigations we will rather work with the Hamiltonian operator $\varPi$ then with the Poisson  brackets which are connected between themself as
\begin{equation}
 {\{} F_1,F_2 {\}} = \textless \nabla F_{1},\varPi \nabla F_{2} \textgreater
\end{equation}
where $\nabla$ denotes the gradient.
Using the last formula we associate the components of the second Hamiltonian operator with the abstract Poisson brackets
\begin{equation}\label{virek}
 {\{} \Phi(x,\theta) , \Phi(x^{'},\theta^{'}) {\}} = \varPi \delta(x-x^{'}) (\theta - \theta^{'} ) =
\left ({\cal D} \partial^{2} + 2 \partial  \Phi(x,\theta)  +  2\Phi(x,\theta)  \partial  + {\cal D} \Phi(x,\theta)  {\cal D} \right ) \delta(x-x^{'}) (\theta - \theta^{'} )
\end{equation}
observing that it is  even  type for the usual $N=1$ supersymmetric
extension of KdV equation while it is odd type
\begin{equation}
 {\{} \Phi(x,\theta) , \Phi(x^{'},\theta^{'}) {\}} = {\cal D}^{-1} \left ( \partial^{3} + 2\partial \Phi_{1}(x,\theta) + 2\Phi_{1}(x,\theta) \partial \right ) {\cal D}^{-1} \delta(x- x^{'}) (\theta -
\theta^{'})
\end{equation}
for  B-extension of  KdV equation.

In the next we will use the following notation: the even or odd bracket denote that the Poisson bracket for
two superfermionic functions $f_1,f_2$ is superbosonic or superfermionic operator,  respectively.
On the other side these brackets could be associated with some algebra.
In the case of the usual KdV equation these brackets are connected with the Virasoro algebra \cite{Gerwazy} while in
the supersymmetric case with the $N=1,2$  supersymmetric Virasoro algebra \cite{Mat2}
 and  for $B$ extension with the $N=1,2$ odd extension of the Virasoro algebra {\cite{Bek,Pop2}}.

\section{ Supersymmetric extension of the Sawada - Kotera equation.}

 In order to generate the Sawada - Kotera equation from the Lax representation let us assume that we have
 the general third order Lax operator of the form
\begin{equation}\label{lax}
 L=\partial^{3} + u \partial  + \lambda u_x
\end{equation}
where $\lambda$ is an arbitrary constant and $u$ is a function of $x,t$.

This operator generates the whole hierarchy of equations,
where the first equation is simply $u_{t}=u_{x}$
and the  nontrivial ones  start from the fifth flow
\begin{equation}\label{rlax}
L_t = 9[L, (L^{\frac{5}{3}})_{+}],
\end{equation}
\begin{equation}
 u_t = ( u_{4x} + 5u u_{xx} + 15 \lambda ( 1-  \lambda ) u_{x}^{2} +
\frac{5}{3}u^3)_x
\end{equation}
only when $ \lambda = \frac{1}{2} , 1, 0$.

For $\lambda λ = \frac{1}{2} $ , we have the Kaup - Kupershmidt hierarchy {\cite{Fad,Fucha}}
while for $\lambda  = 0,1$
we obtain the Sawada - Kotera hierarchy {\cite{Fad,Fucha}}

The Kaup-Kupershmidt and Sawada-Kotera equations have a similar  Bi-Hamiltonian structure where the  symplectic operator $\Gamma$ generates
\begin{eqnarray}
u_t & = & \varGamma \frac{\delta H}{\delta u} = ( \partial^{3} + \gamma (\partial_x u + u \partial_x) )
\frac{\delta H}{\delta u} \\ \
H & = & \frac{1}{2} \int dx ( uu_{xx}  + \beta u^3)
\end{eqnarray}
where   $ \gamma = \frac{1}{2}$ and $\beta=\frac{4}{3} $ for  Kaup - Kupershmidt
equation while for  Sawada - Kotera equation we have
$\gamma = 2$ and $\beta=\frac{1}{3}$.

The implectic operator for these equations could be obtained
from the factorization of the recursion operator \cite{Fucha}
\begin{equation}
 \frac{\delta H_{1}}{\delta u}  =\Omega u_{t}= (\partial^{2} + k_{1}(\partial u + u\partial ) +
\partial^{-1}(k_{1}u_{xx} + k_{3}u^{2})  + (k_{2}u_{xx}+ k_{3}u^{2}) \partial^{-1}) u_t
\end{equation}
where $ k_{1}=\frac{5}{2},k_{2}=1,k_{3}=2$ for the Kaup - Kupershmidt equation while for the Sawada-Kotera
equation  $ k_1=k_3=1,k_3=\frac{1}{2}$.

Let us now consider in details the supersymmetric $N=1$ version of the Lax operator (\ref{lax}). The most general
form of such   generalization is
\begin{equation}\label{slax}
 L=\partial^{3}  + \lambda_{1} \Phi_{1}\partial   +
\lambda_{2} \Phi_{x}{\cal D} + \lambda_{3} \Phi_{1,x}
\end{equation}
where $\lambda_{i}$ are  the arbitrary constants and $\Phi$ is the superfermionic field of the
fractional conformal dimension $\frac{3}{2}$ with the decomposition $\Phi=\xi + \theta w$.

The Lax representation  (\ref{rlax})  gives us the consistent equation only for the special choice of the
parameters $\lambda_{i}$. We have five solutions for the coefficients.

1.) For $\lambda_{1}=2,\lambda_2=0,\lambda_3=1$  we obtain  B-extension of Kaup - Kupershmidt
equation
\begin{equation}
 \Phi_{t}=\Phi_{5x} + 10 \Phi_{xxx} \Phi_{1} +
15\Phi_{xx}\Phi_{1,x} + 10\Phi_{x}\Phi_{1,xx} + 20 \Phi_{x}\Phi_{1}^{2}
\end{equation}

2.) We have  B-extension of  Sawada - Kotera equation

\begin{equation}
 \Phi_{t}=\Phi_{5x} + 5 \Phi_{xxx} \Phi_{1} +
5\Phi_{x}\Phi_{1,xx} + 5 \Phi_{x}\Phi_{1}^{2}
\end{equation}

for the
$ \lambda_{1}=\lambda_{3} =1,\lambda_{2}=0$ or $ \lambda_1=\lambda_2=1,\lambda_3=0$ or
$ \lambda_2=\lambda_3=0,\lambda_1=1$.
\vspace{0.5cm}

3.) We have  $N=1$ supersymmetric extension of  Sawada - Kotera equation considered in \cite{Liu}
\begin{equation}
 \Phi_{t}=\Phi_{5x} + 5 \Phi_{xxx}\Phi_{1} +
5\Phi_{xx}\Phi_{1,x} + 5 \Phi_{x}\Phi_{1}^{2}
\end{equation}

for $\lambda_1=\lambda_3=1,\lambda_2=-1$. In the components last equation reads
\begin{eqnarray}{}
 \xi_{t} &=& \xi_{5x} + 5w\xi_{xxx} + 5w_{x}\xi_{xx} + 5u^{2}\xi_{x} \\ \
 w_{t} &=& w_{5x} + 5ww_{xxx} + 5w_{x}w_{xx} +5w^{2}w_{x} - 5\xi_{xxx}\xi_{x}
\end{eqnarray}

from which we see that the bosonic sector is modified by the term $\xi_{xxx}\xi_{x}$ and therefore it
is not a $B$ extension.

\section{Odd Bi-Hamiltonian structure.}

For the B-Extension of the Kaup - Kupershmidt or the Sawada - Kotera equations  it is easy to construct  conserved
currents and a Bi-Hamiltonian structure using the prescription  described above.
Hence these generalizations are integrable. For the supersymmetric  $N=1$ generalization of  the Sawada - Kotera
equation the situation is much more complicated. In order to find the Bi - Hamiltonian structure let us first make several observations.

\textbf{I} :  The Lax operator (\ref{slax}) does not produce the supersymmetric generalization of the
Kaup - Kupershmidt equation. Taking into account that the classical Kaup - Kupershmidt  equation, as well as the
classical Sawada - Kotera equation,  possess  the same Hamiltonian operator (see Esq. 31 ),
we tried to apply the supersymmetrized version of this operator (\ref{virek})  in order to obtain supersymmetric generalization
of these equations.
We verified that this approach gives us some supersymmetric equations whose   bosonic limit  is reduced  to the
Sawada - Kotera or to the Kaup - Kupershmidt equation but these equations does not possess the higher order
conserved quantities and hence they are not integrable.

 \textbf{II}:  The Lax operator (\ref{slax}), as it was shown in {\cite{Liu}},  could  be factorized as
\begin{equation}
 L=\partial_{xxx} +\Phi_{1}\partial -\Phi_{x}{\cal D} +\Phi_{1,x} = \varLambda^{2} =
\left ({\cal D}^3 + \Phi \right )^{2}
\end{equation}
The $\varLambda$ operator belongs to the reduced  Manin-Radul supersymmetric KP hierarchy
\cite{Aratyn1}.
Due to this factorization we obtain new hierarchy of the supersymmetric equations
\begin{equation}
 \varLambda_{t,k}= 9 [\varLambda , (\varLambda)^{\frac{k}{3}}_{+} ]
\end{equation}
where $k$ is a natural number such that $k \neq 3n, 4n , 4n + 1, n=0,1,\dots$.

Let us present the first four  equations
\begin{equation}\label{phi2}
 \Phi_{t,2} =\Phi_{x},
\end{equation}
\begin{equation}\label{phi7}
  \Phi_{\tau,7} = ( \Phi_{1,xx} + \frac{1}{2}\Phi_{1}^2 +3\Phi\Phi_{x} )_{x}
\end{equation}
\begin{equation}\label{phi10}
 \Phi_{t,10}=\Phi_{5x} + 5 \Phi_{xxx}\Phi_{1} +
5\Phi_{xx}\Phi_{1,x} + 5 \Phi_{x}\Phi_{1}^{2}
\end{equation}
\begin{eqnarray}\label{phi11}
 \Phi_{\tau,11} &=& \Phi_{1,5x} + 3\Phi_{1,xxx}\Phi_{1} + 6\Phi_{1,xx}\Phi_{1,x} +
2\Phi_{1,x} \Phi_{1}^{2} \\ \nonumber
 && -3\Phi_{4x}\Phi  - 2\Phi_{xxx}\Phi_{x} -6\Phi_{xx}\Phi\Phi_{1} -6\Phi_{x}\Phi\Phi_{1,x}
\end{eqnarray}
where $t$ denotes usual even time while $\tau$ odd time.

Moreover this factorization implies that the usual formula on
the conserved quantities, as the residue of $L^{\frac{n}{3}}$ does not lead
us to any conserved quantities \cite{Liu} . On the other side if we apply this formula to
 $\varLambda$ operator,  we obtain the whole hierarchy of local superfermionic conserved quantities.
 For example the first three  quantities are
\begin{eqnarray}\label{h1}
H_1 & = & \int\, dx d\theta\,  \Phi\Phi_x = \int \, dx\,\, \xi_{x}w \\  \label{h2}
H_2 &=& \frac{1}{6}\int \,dx d\theta\, \left (3\Phi\Phi_{xxx} + 2\Phi_{1}^3 \right ) = \int \, dx\,  \left (
\xi_{xxx}w  + \xi_{x}w^2 \right )\\ \label{h3}
H_3 &=& \int \, dx d\theta \, \left( \Phi_{7x}\Phi + 8\Phi_{xxx}\Phi\Phi_{1,xx} +
\Phi_{x}\Phi (4\Phi_{1,4x} + 20\Phi_{1,xx}\Phi_{1} + 10 \Phi_{1,x}^{2} + \frac{8}{3}\Phi_{1}^{3}) \right )
 \end{eqnarray}

These charges are conserved  for the supersymmetric Sawada - Kotera equation and
does not reduce in the bosonic limit, where all fermions
functions disapeare, to the classical charges.
Therefore in general,  we can not conclude that the integrability of the supersymmetric models
implies the integrability of  bosonic sector.

\
\textbf{III}:  Let us  notice that the exotic equations  (\ref{phi7}) , (\ref{phi11})  and  $\Phi_{\tau,19}$ (as we also checked )
  could be rewritten as
\begin{eqnarray}\label{phi77}
 \Phi_{\tau ,7}  = &&\varPi\Phi_{t,2} =  \left ({\cal D} \partial^{2} + 2 \partial  \Phi  +  2\Phi  \partial  + {\cal D} \Phi {\cal D}  \right ) \frac{\delta H_1}{\delta \Phi} \\ \
&& \Phi_{\tau ,11} =  \varPi \frac{\delta H_2}{\delta \Phi}, \hspace{1.5cm}
 \Phi_{\tau,19} =\varPi \frac{\delta H_3}{\delta \Phi}
\end{eqnarray}
and hence the usual even supersymmetric second Hamiltonian operator of the supersymmetric Korteweg - de Vries equation
 see Esq. \ref{virek} creates the exotics equations. However this Hamiltonian operator is not our desired
Hamiltonian operator which generates  the physical equations.

\textbf{IV}: The densities of the conserved quantities (\ref{h1}) - (\ref{h3})  are  superbosonic functions
and hence their gradients are  superfermionic functions.
As $\Phi$ is a superfermionic  function,  it forces that the expected Hamiltonian operator, responsible for the creation of physical equations, to be a
superboson operator. This conclusion implies that such Hamiltonian operator, if exists,   creates an  odd
Hamiltonian structures with the antibrackets as the Poisson brackets.

This observations allow us to construct an odd  Bihamiltonian structure for the $N=1$ supersymmetric Sawada - Kotera as follows:

First let us notice that the supersymmetric $N=1$ Sawada - Kotera equation could be rewritten as

\begin{equation}
 \Phi_{t,10} = \left  ( {\cal D} \partial^{2} + 2 \partial  \Phi  +  2\Phi \partial  + {\cal D} \Phi  {\cal D}
\right ) \left ( \Phi_{1,xx} + \frac{1}{2} \Phi_{1}^{2} + 3\Phi\Phi_{x} \right )
\end{equation}

In this formula we have  correct  Hamiltonian operator, while the last term is not a gradient of some
superfunction. However if we differentatiate this  term,  then it becomes  seventh flow in our hierarchy
(\ref{phi7}). So using the formula (\ref{phi77})  we arrive to the following theorem
\vspace{0.3cm}

\textbf{Theorem 1}: The operator
\begin{equation}\label{odek}
 P = \left  ( {\cal D} \partial^{2} + 2 \partial  \Phi  +  2\Phi \partial  + {\cal D} \Phi  {\cal D}\right )
 \partial^{-1}   \left  ( {\cal D} \partial^{2} + 2 \partial  \Phi  +  2\Phi \partial  + {\cal D} \Phi
 {\cal D} \right ) =\varPi \partial^{-1}\varPi
\end{equation}
is a proper Hamiltonian operator which generates the supersymmetric $N=1$ Sawada - Kotera equation
$\Phi_{t,10}  = P\frac{\delta H_1}{\delta \Phi}$
and satisfy the Jacobi identity.
\vspace{0.4cm}

\textbf{Proof}. We have to check the following identity \cite{Blacha}
\begin{equation}
\textless \alpha , P^{'}_{P\beta} \gamma  \textgreater\quad +\quad \textless \beta , P^{'}_{P\gamma} \alpha  \textgreater \quad + \quad \textless \gamma , P^{'}_{P\alpha} \beta  \textgreater = 0.
\end{equation}
where
$P^{'}_{P\beta}$ denotes the
Gatoux derivative along the vector $ P\beta$ and $\alpha,\beta,\gamma$ are the test superfermionic
functions. After a lengthly \footnote{the computations are simplified if one use the computer algebra
Reduce \cite{Reduce}  and the package Susy2 \cite{Pop3} } calculation these formula could be reduced to
the form in which the typical term is
\begin{equation}
 \int dx d\theta \Big (J_{0} + J_{1,1}\partial^{-1}J_{1,2} + J_{2,1}\partial^{-1}J_{2,2} \partial^{-1}J_{2,3} +
 ....\Big )
\end{equation}
where the expressions $J_0,J_{n,m} $ are constructed  out of $\Phi,\alpha,\beta,\gamma$ and their different (susy)derivatives but does not contain the integral operator. Using the rule
\begin{equation}
 \alpha_{1} = \overleftarrow{{\cal D}} \alpha + \alpha \overrightarrow{{\cal D}} \quad  ,\quad
\alpha_{x} = \overleftarrow{\partial} \alpha - \alpha \overrightarrow{\partial}
\end{equation}
we can eliminate (susy)derivative from the
test superfermionic function $\alpha$ then from $\beta$ and finally from $\gamma$. As the result
 we obtain that   the Jacobi identity reduces to the form with  the typical terms
\begin{equation}
  \int dx d\theta \Big ( \Phi_{n}\alpha_{m}
\Phi_{k}\beta_{l} \,\, {\bf \partial^{-1}} \,\, \Phi_{s}\gamma_{r} \,\,  \pm \,\,  \Phi_{s}\gamma_{r}\,\,\, {\bf\partial^{-1}}  \,\,
 \Phi_{n}\alpha_{m}  \Phi_{k} \beta_{l}  +
\Phi_{\hat n}\alpha_{\hat m} \,\,{\bf \partial^{-1}} \,\,
\Phi_{\hat k} \beta_{\hat l} {\bf \,\, \partial^{-1}}\,\,  \Phi_{\hat s} \gamma_{\hat r}+ ...\Big )
\end{equation}
where the  indices can take values $0,1,2,3,4$ and $\pm$ depends on the parity of the
superfunction under the integral operator. Due to the antisymmetric  property of integral operator
$\partial^{-1}$ all terms cancels out and the Jacobi identity holds. $\bullet$

It that way we obtained the odd Hamiltonian operator which generates the physical equations and as well as the
supersymmetric Sawada - Kotera equation.
Regarding $P$ operator as a linear map from the co-vector fields to the vector fields
the odd Hamiltonian operator can now be decomposed into
\begin{equation}
 \frac{d}{d t} \left(\begin{tabular}{cc} $u$ \\
 $\xi$
\end{tabular}\right) =
\left(\begin{tabular}{ccc}

$
\begin{tabular}{c}
$2(\partial^{2} \xi_{x} + \xi_{x} \partial^{2}) + 3 \partial \xi_{x}  \partial +$ \\
$ 2\xi_{x}\partial^{-1} w_{x} - 2w_{x}\partial^{-1} \xi_{x} + 4w\xi_{x}$
\end{tabular}
$ & $
-\left(
 \begin{tabular}{c}
 $ \partial^{4} +2w_{x}\partial^{-1}w + 4w^{2} +\partial^{2} w $ \\
$ +2 (\partial w + w\partial)\partial - 2\xi_{x} \partial^{-1} \xi_{x} $
\end{tabular}\right)
$
\\
$
\begin{tabular}{c}
 $ \partial^{4} -2w\partial^{-1}w_{x} + 4w^{2} + $ \\
$w\partial^{2} + 2\partial(\partial w + w\partial) - 2\xi_{x} \partial^{-1} \xi_{x} $
 \end{tabular}
$ & $
\begin{tabular}{c}
$ \xi_{x} \partial +\partial\xi_{x}    $ \\
$ - 2\xi_{x}\partial^{-1} w -2w\partial^{-1}\xi_{x}$
\end{tabular}
$
\end{tabular}\right)
\left (
\begin{tabular}{cc}
$\frac{\delta H}{\delta u}$ \\
$$\\
$ \frac{\delta H}{\delta \xi} $
       \end{tabular}\right)
\end{equation}
\vspace{0.3cm}

We tried  to verify the validity of the accidentally found Hamiltonian operator  (\ref{odek}) using
a natural way of associating several Hamiltonians structure to a given Lax operator {\cite{Hamy1}} -{\cite{ Hamy6}} .
 This approach yields multi - Hamiltonian formulations for the isospectral flows connected to the scattering problem
  given by that Lax operator. Let us briefly presents the main steps of this procedure. The multi- Hamiltonian structure
   $\Gamma$ could be recovered from the given Lax operator computing

\begin{eqnarray}
\dot{L} =  \Gamma_{1} \nabla F & = & (L (\nabla F))_{+} - ((\nabla F) L)_{+} \\ \
\dot{L} = \Gamma_{2} \nabla F  & =  & L((\nabla F)L)_{+} - (L(\nabla F))_{+}L \\ \
\dot{L} = \Gamma_{3} \nabla F  & =  & L (L(\nabla F) L)_{+}- (L(\nabla F)L)_{+}L -
L((\nabla F)L)_{+}L + L(L(\nabla F))_{+}L
\end{eqnarray}
where $\nabla F$ denotes the gradient of some conserved quantity.

Usually the multi - Hamiltonians
 $\Gamma_{i}$ are called the ``linear``, the ''quadratic`` and  the ''cubic`` operators for $i=1,2,3$, respectively.
Given a Hamiltonian function $H(L)$ ,where the Lax operator $L$ may be regarded as element of the algebra of super pseudo-differential operators
\begin{equation}
 L:=\sum_{k<\infty} (a_k + b_k{\cal D}) \partial^{k}
\end{equation}
a convenient parametriaztion of gradient $\nabla H$ is
\begin{equation}
 \nabla H =  \sum_{k \geq 0}  \partial^{-k-1} ( -{\cal D} \frac{\delta H}{\delta a_k} + \frac{\delta H}{\delta b_k} )
\end{equation}
In this parametrisation the trace duality has the  usual Euclidean form.

Now, trying to evaluate the first Hamiltonian operator $\varGamma_1$ for $L={\cal D}\partial + \Phi$ , one immediately encounters a technical difficulty: the corresponding Poisson bracket cannot be properly restricted
to this Lax operator. Therefore one should  first embed this operator into a larger subspace as
\begin{equation}
 L={\cal D}\partial + a\partial +\Phi + b{\cal D}
\end{equation}
and assume that
\begin{equation}
 \nabla H  = \partial^{-1} ( \frac{\delta H}{\delta b} - {\cal D}\frac{\delta H}{\delta \Phi}) -
 \partial^{-2}{\cal D}  \frac{\delta H}{\delta a}
\end{equation}

Thus the Hamiltonian equation $\dot{L} =\Gamma_{1} \nabla H $ could be transformed to the matrix form
\begin{equation}
 \left (\begin{tabular}{c}
$ \dot{b} $ \\  $$  \\
    $ \dot{\Phi} $ \\ $$  \\$ \dot{a}$
\end{tabular}\right )
=
\left( \begin{tabular}{ccc}\label{jeden}
$ 0$ & $ -{\cal D}$  & $ 0 $ \\
$$ & $$ & $$ \\
$ - {\cal D} $ & $ 0 $ & $ 2 $  \\
$$ & $$ & $$ \\
$ 0 $ & $ 2 $ & $ 0 $
\end{tabular} \right)
\left(\begin{tabular}{c}
$ \frac{\delta H}{\delta b}$ \\ $ $ \\
    $ \frac{\delta H}{\delta \Phi}$ \\ $ $\\
        $ \frac{\delta H}{\delta a} $
      \end{tabular} \right )
\end{equation}
The next step is to restrict this operator to the smaller subspace where  our Lax operator lives. It means
that we should apply the Dirac reduction technique to the subspace where $ a=0$ and $b=0$. We have the standard reduction
lemma for Poisson brackets {\cite{Hamy6}} which can be formulated as follows.

For the given Poisson tensor
\begin{equation}
 P(v,w) =\left(\begin{tabular}{cc}
$ P_{v,v} $ & $ P_{v,w} $ \\
$ P_{w,v} $ & $ P_{w,w} $
\end{tabular}\right )
\end{equation}
let us assume that $P_{vv}$ is invertible, then for arbitrary $v$ the map given by
\begin{equation}
 \varTheta(w:v) = P_{w,w} - P_{w,v}P_{v,v}^{-1}P_{v,w}
\end{equation}
is a Poisson tensor where $v$ enters as a parameter rather than as a variable.

Unfortunately it is impossible to make such Dirac reduction for the matrix (\ref{jeden}). The same situation
occurs  in the classical case.

The situation changes when we compute quadratic brackets. Then the analogous matrix to the matrix in (\ref{jeden}) is

\begin{equation}
 \left (\begin{tabular}{ccc}
$ {\cal D}\partial + 2\Phi $ & $ -\Phi {\cal D} $ & $ -\partial $ \\
$ -{\cal D}\Phi $ & $ -{\cal D}\partial^{2} - \partial \Phi - \Phi \partial $  & $ {\cal D}\partial $ \\
$ \partial  $ & $ -{\cal D}\partial  $ & $ {\cal D} $
\end{tabular}\right )
\end{equation}
Now we can make the Dirac reduction with respect to the last column and last row and  we obtain

\begin{equation}
{\hat{\varGamma_{1}}} =
 \left (\begin{tabular}{cc}
$ 2({\cal D}\partial + \Phi) $ & $ -({\cal D}\partial +\Phi ) {\cal D} $ \\
$ -{\cal D} ( {\cal D}\partial +\Phi ) $ & $ -\partial \Phi - \Phi \partial $
\end{tabular}\right )
\end{equation}
Noticing that
\begin{equation}
 {\hat{\varGamma}}_{2,1}{\hat{\varGamma}}_{1,1}^{-1} {\hat{\varGamma}_{1,2}} =  \frac{1}{2} (
\partial^{2}{\cal D} + {\cal D}\Phi {\cal D} )
\end{equation}
we can finally carry out  the reduction with respect to the first column and the first row obtaining that the
second Hamiltonian operator is proportional to   second Hamiltonian operator of the supersymmetric
Korteweg de Vries equation $\varGamma_{1} = -\frac{1}{2} \varPi$.

The third Hamiltonian operator produces the complicated 6 $\times$  6 matrix

\begin{equation} \label{trzy}
{\hat\varGamma} _{3} =
 \left( \begin{tabular}{cccccc}
$ \partial \Phi_{1} - \Phi_{x} {\cal D}  $ &
    $-\partial \Phi_{1} {\cal D}   $ &
        $ 0 $ &
            $ \Phi_{x} $ &
                $  \partial $ &
                    $  -\partial {\cal D}  $  \\
$ {\cal D} \Phi_{1} \partial $ & $ -\partial \Phi_{1} \partial $ & $  - \partial^{2} {\cal D} - {\cal D}\Phi_{1} $ & $  \Phi_{1,x} $ & $  \partial {\cal D} $ & $ \Phi_{1} $ \\
$ 0 $ &
    $  -{\cal D}\partial^{2} - \Phi_{1}{\cal D} $ &
        $ 2\partial $ &
            $ {\cal D}\partial $ &
                $ 0 $ & $ -{\cal D}$ \\
$ -\Phi_{x} $ &
    $ \Phi_{1,x} $ &
        $ -{\cal D}\partial $ &
            $ \Phi_{1} $ &
                $ {\cal D} $ & $ \partial $ \\
$ \partial $ &
        $ -{\cal D}\partial $ &
            $ 0 $ &
                $ {\cal D} $ &
                    $ 0 $ &
                        $ 0 $ \\
$ {\cal D}\partial $ & $ \Phi_{1} $ & $ -{\cal D} $ & $ - \partial$ & $ 0 $ & $ 2 $
\end{tabular} \right )
\end{equation}

Again we can invoke Dirac reduction to the subspace spanned by $\Phi$ only. We first make the reduction
with respect to the fourth, fifth and sixth row and fourth, fifth and sixth column. Taking into account that
it is possible to find the inverse matrix to the matrix constructed out of these columns and rows
\begin{equation}
 \left( \begin{tabular}{ccc}
$ \Phi_{1} $ & $ {\cal D} $ & $ \partial $ \\
$ {\cal D} $ & $ 0 $ & $ 0 $ \\
$-\partial $ & $ 0 $ & $ 2 $
\end{tabular}
\right )^{-1}  = \left ( \begin{tabular}{ccc}
$ 0 $ & $ {\cal D}\partial^{-1} $ & $ 0 $ \\
$ {\cal D} \partial^{-1} $ & $ -{\cal D} \partial^{-1} \Phi - \Phi {\cal D}\partial^{-1} - \frac{1}{2}\partial $
& $ -\frac{1}{2}{\cal D} $ \\
$ 0 $ & $\frac{1}{2} {\cal D} $ & $ \frac{1}{2} $
                       \end{tabular}\right )
\end{equation}

we found  the reduced matrix in the form
\begin{equation}
\widetilde{\varGamma_{3}} =
 \left( \begin{tabular}{ccc}
$ 2\partial^{3} - 2\Phi_{x} {\cal D} + 2\partial \Phi_{1} $ &
$ -\partial^{3}{\cal D} + \Phi_{x}\partial - \partial \Phi_{1} {\cal D} $ & $ 0 $ \\
$ \partial^{3} {\cal D} + \partial \Phi_{x} + {\cal D} \Phi_{1} \partial $ &
     $ -\frac{1}{2} ( \partial^{4} + 2\partial \Phi_{1}\partial + \Phi_{1,xx} + \Phi_{1}^{2})$ &
            $ -\frac{1}{2} ( \partial^{2}{\cal D} + {\cal D} \Phi_{1} +\Phi_{x} )  $ \\
$ 0 $ &
    $ \frac{1}{2}(\partial^{2}{\cal D} + \Phi_{1}{\cal D} - \Phi_{x})  $ &
        $ \frac{3}{2}\partial  $
\end{tabular} \right)
\end{equation}

Again we carry out the Dirac reduction for  the last column and last row  in $\widetilde{\varGamma_{3}}$ obtaining

\begin{equation}
\widehat{\varGamma_{3}} =
 \left( \begin{tabular}{cc}
$ 2( \partial^{3} - \Phi_{x} {\cal D} + \partial \Phi_{1}) $ &
$ -(\partial^{3} - \Phi_{x}{\cal D} + \partial \Phi_{1})  {\cal D} $  \\
$ {\cal D} ( \partial^{3}  -  \Phi_{x} {\cal D} + \partial  \Phi_{1})  $ &
     $ \widehat{\varGamma}_{3,2,2} $
\end{tabular} \right)
\end{equation}

where
\begin{equation}
 \widehat{\varGamma}_{3,2,2} = -\frac{1}{3} ( 2\partial^{4}  + 4\Phi_{1}\partial^{2} + 3\Phi_{1,x}\partial  +2\Phi_{x}{\cal D}\partial + \Phi_{1,xx} + 2\Phi_{1}^{2} + \Phi_{xx}{\cal D} -
{\cal D} \Phi_{1}\partial^{-1}\Phi_{x} + \Phi_{x}\partial^{-1}\Phi_{1}{\cal D})
\end{equation}

It is possible to carry out the last reduction for the first column and first row in $
\widehat{\varGamma_{3}}$ because
\begin{equation}
 \widehat{\varGamma_{3}}_{2,1}\widehat{\varGamma_{3}}_{1,1}^{-1} \widehat{\varGamma_{3}}_{1,2} =
\frac{1}{2} {\cal D} ( \partial_{xxx} - \Phi_{x}{\cal D}  + \partial\Phi_{1}) {\cal D}
\end{equation}
and as the result we obtained that the  third Hamiltonian operator is proportional to our odd Hamiltonian operator
(\ref{odek})   $ \varGamma_{3} = -\frac{1}{6} P$
\vspace{0.3cm}

As we seen these two Hamiltonian operators produces  two different series of equations, the even $\varGamma_{2}$ operator generates the exotic equations, while the odd one $\varGamma_{3}$ generates  the supersymmetric Sawada - Kotera equation.
So these operators could be considered independently and
one can ask wheather  these two series possess  second  Hamiltonian operator. For the  odd Hamiltonian operator
we can find such by factorization of the recursion operator found in \cite{Liu} as
 $R = P J $. As the result we were able to prove  the following theorem

\textbf{Theorem 2}: The $J$  operator defines  a proper implectic operator for the
$N=1$ supersymmetric Sawada - Kotera equation
\begin{equation}
 J = \partial_{xx} +\Phi_{1} - \partial^{-1} \Phi_{1,x} + \partial^{-1}\Phi_{x} {\cal D} +
\Phi_{x}\partial^{-1}{\cal D}
\end{equation}

\textbf{Proof}. The  implectic operator should satisfy {\cite{Blacha}}
\begin{equation}
\textless \alpha , P^{'}_{\beta} \gamma  \textgreater\quad +\quad \textless \beta , P^{'}_{\gamma} \alpha  \textgreater \quad + \quad \textless \gamma , P^{'}_{\alpha} \beta  \textgreater = 0.
\end{equation}
After computing this expression we follow the same strategy as in the theorem 1
and verify that this is indeed zero. $\bullet$

This operator generate the gradient of the conserved quantity according to the formula
\begin{equation}
 J\Phi_t=\frac{\delta H_3}{\delta \Phi}
\end{equation}

On the other side the implectic operator could be decomposed in the following manner
\begin{equation}
\left(
 \begin{tabular}{c}
$\frac{\delta H}{\delta w} $ \\
$$ \\
$ \frac{\delta H}{\delta \xi}$
\end{tabular} \right )=
\left(
\begin{tabular}{ccc}
 $ \xi_{x}\partial^{-1} + \partial^{-1}\xi_{x} $ &
$ $ &
$ \partial^{2} -\partial^{-1} w_{x} + w $ \\
$ $ & $ $ & $ $  \\
 $-\partial^{2} - w_{x}\partial^{-1} - w $ & $ $ &
$ \xi_{x} $ \end{tabular}
\right )
\left(
\begin{tabular}{c}
$ w_{t} $ \\
$$ \\
$ \xi_{t} $
\end{tabular}
\right)
\end{equation}

The recursion operator found in \cite{Liu} does not generate  exotic flows in   the
 supersymmetric  Sawada - Kotera hierarchy. More precisely  we checked that $ R\Phi_{\tau,7} \neq \Phi_{\tau,14}$.
 Unfortunately we were not been able to find any second Hamiltonian  or recursion operator for  the exotic series.

\section{Conclusion}

In this paper we found new  unusual features of the supersymmetric models,
We showed that the supersymmetric  extension of the Sawada - Kotera equation has  an odd Bi-Hamiltonian structure. The exotic equations in the supersymmetreic Sawada- Kotera hierarchy are generated by the same supersymmetric Hamiltonian operator which appeared in the supersymmetric Korteweg de Vries equation.
The existence of the Bi-Hamiltonian structure allow us to state that this model is integrable.
Unfortunately we did not found  any second Hamiltonian operator or recursion operator responsible for
 exotics equations. It seems reasonable to assume that the second recursion operator should exist
 which is supported by the
observation that in the classical models, without supersymmetry,  two  different recursion operators
could exists \cite{recydywa}, \cite{recydywa1}.
It will be also interesting to find more examples of the   supersymmetric models
 with the  similar or the same properties as  the supersymmetric Sawada - Kotera equation.

\end{document}